\documentstyle[11pt]{article}

\newcommand{\beq}{\begin{equation}}
\newcommand{\eeq}{\end{equation}}
\newcommand{\p}{\partial}
\newcommand{\bfe}{\mbox{\bf e}}

\newcommand{\disp}{\displaystyle}
\newcommand{\no}{\nonumber}
\newcommand{\U}{{\cal U}}
\renewcommand{\H}{{\cal H}}
\renewcommand{\L}{{\cal L}}

\begin{document}

\begin{center}

{\Large{\bf Note on ``Electromagnetism and Gravitation''}}\\

\vspace{.5in}

{\large {\bf Kenneth Dalton}} \\

\vspace{.3in}

e-mail: kldalton@cs.clemson.edu \\

\vspace{1.in}

{\bf Abstract }

\end{center}

\vspace{.25in}

We obtain Hamilton equations for the gravitational field and demonstrate
the conservation of total energy.  We derive the Poisson bracket equation
for a general dynamical variable.

\clearpage

\section*{\large {\bf 1. Lagrange Equations}}

\indent

The gravitational Lagrangian is given by [1]
\begin{eqnarray}
  L_G &=& \frac{c^4}{8\pi G}
           g^{\alpha\beta}Q^\rho_{[\eta\alpha]}Q^\eta_{[\rho\beta]}\no\\
      &=& \frac{c^4}{8\pi G}
        \left\{g^{00}Q^l_{m0}Q^m_{l0} + g^{lm}Q^0_{0l}Q^0_{0m}\right\}\no\\                               
      &=& \frac{c^4}{32\pi G}
           \left\{g^{00}g^{la}g^{mb}
            \frac{\p g_{am}}{\p x^0}\frac{\p g_{bl}}{\p x^0}
           + g^{lm}g^{00}g^{00}
              \frac{\p g_{00}}{\p x^l}\frac{\p g_{00}}{\p x^m}\right\}                             
\end{eqnarray}
$Q^\mu_{[\nu\lambda]}$ is the gravitational field tensor, with non-zero 
components

\beq
   Q^i_{[j0]} = Q^i_{j0} = \frac{1}{2} g^{ia}\frac{\p g_{aj}}{\p x^0}                         
\eeq
\beq
   Q^0_{[0i]} = Q^0_{0i} = \frac{1}{2} g^{00}\frac{\p g_{00}}{\p x^i}
\eeq
Field equations were derived in [1] under the assumption that the seven
potentials $g_{\mu\nu} = (g_{00}, g_{ij})$ are independent.  However, 
$L_G$ does not contain the time derivative of $g_{00}$ and, therefore,
it cannot be a true dynamical variable.  In this note, we will eliminate
$g_{00}$ from the Lagrangian, in order to establish Hamilton equations of
motion.  This is accomplished via the principle of {\it space-time 
reciprocity}.

According to Einstein, an observer at rest in a gravitational field is 
equivalent to an accelerated observer in free space.  Moreover, the 
difference in gravitational potential between two points P and P' is 
equivalent to a relative velocity between observers at P and P'  [2].
It follows that:
\begin{itemize}
 \item[(a)] time intervals measured at P and P' are related by \newline 
 $\Delta t =\Delta t'/\sqrt{1-v^2/c^2}$\hspace{.2in} (time dilatation);
 \item[(b)] distance intervals measured at P and P' are
    related by \newline
 $\Delta l = \Delta l'\sqrt{1 - v^2/c^2}$\hspace{.2in} (length contraction).
\end{itemize}
The reciprocity in space and time gives way to the equality
$ \Delta t \, \Delta l = \Delta t'\, \Delta l'$.  We state the more general 
principle as follows: the space-time volume element $\sqrt{-g}\, d^4 x $
is not affected by the presence of a gravitational field.

The array of potentials $g_{\mu\nu}$ always takes the form

\beq
         g_{\mu\nu} = \left( \begin{array}{cccc}
                                          g_{00}&0&0&0 \\
                                          0&&&  \\
                                          0&&g_{ij}&  \\
                                          0&&&
                                          \end{array}
                      \right)
\eeq
Setting $\mbox{\rm det}\, g_{\mu\nu} =-g $ and 
$\mbox{\rm det}\, g_{ij}= -h $, we have

\beq
      \sqrt{-g} = \sqrt{g_{00}}\, \sqrt{-h}
\eeq
However, by space-time reciprocity, the density $\sqrt{-g}$ is equal to
the associated flat-space density $\sqrt{-h_0}$.  For example, in 
rectangular coordinates, $\sqrt{-g} = 1$; in spherical coordinates, 
$\sqrt{-g} = r^2 sin{\theta}$; etc. Thus,

\beq
      \sqrt{-g} = \sqrt{g_{00}}\, \sqrt{-h} = \sqrt{-h_0}
\eeq
or
\beq    
     g_{00} = \frac{h_0}{h}
\eeq
This constraint serves to eliminate $g_{00}$ from the Lagrangian.

We now derive the corresponding field equations.  Since $\sqrt{-g} =
\sqrt{-h_0}$ does not depend upon the gravitational field, its variation
is zero

\beq
  \delta\sqrt{-g} = - \frac{1}{2}\sqrt{-g}\, g_{\mu\nu}\delta g^{\mu\nu} = 0  
\eeq              
or
\beq
     g_{00} \delta g^{00} = -g_{ij} \delta g^{ij}
\eeq
This shows that, at any point, $\delta g^{00}$ is determined by the 
$\delta g^{ij}$. Variation and integration by parts yields

\begin{eqnarray}
&& \hspace{-.3in} \delta \int L_G \sqrt{-g}\, d^4 x = \no \\ 
&& =  \int \frac{c^4}{8\pi G}\left\{\frac{\p}{\p x^0}
      (\sqrt{-g}\,g^{00}Q^i_{j0}) - \delta^i_j \frac{\p}{\p x^l}
      (\sqrt{-g}\,g^{lm}Q^0_{0m})\right\}g_{il} \delta g^{lj}\, d^4 x \no \\
&& \hspace{.5in} +\, \frac{1}{2} \int \sqrt{g^{00}}(T^{\,\,\,i}_{j\,\,G}
      -\delta^i_j T^{\,\,\,0}_{0\,\,G})g_{il} \delta g^{lj} \sqrt{-g}\,d^4 x
\end{eqnarray}
The gravitational stress-energy-momentum tensor is 

\beq
    T^{\mu\nu}_{\,\,G} = \frac{c^4}{4\pi G} \sqrt{g_{00}} \left\{
     g^{\mu\alpha}g^{\nu\beta}Q^\rho_{[\eta\alpha]}Q^\eta_{[\rho\beta]}
     - \frac{1}{2} g^{\mu\nu}g^{\alpha\beta}Q^\rho_{[\eta\alpha]}
       Q^\eta_{[\rho\beta]} \right\}
\eeq
and it is understood that $g_{00} = h_0/h $.  The contributions of matter
and electromagnetism are expressed by

\beq
 \delta \int L_M \sqrt{-g} \, d^4 x = \frac{1}{2} \int \sqrt{g^{00}}
 (T^{\,\,\,i}_{j\,\,M}-\delta^i_j T^{\,\,\,0}_{0\,\,M}) g_{il}\delta g^{lj} 
      \sqrt{-g}\, d^4 x 
\eeq
where

\beq
 T^{\mu\nu}_{\,\,M} = \sqrt{g_{00}} \left\{\rho c^2 u^\mu u^\nu 
 + F^\mu_{\,\,\,\alpha}F^{\alpha\nu} + \frac{1}{4} g^{\mu\nu}
   F_{\alpha\beta} F^{\alpha\beta} \right\}
\eeq
(The factor of $\sqrt{g_{00}}$ in (11) and (13) is discussed in section 3.)
Combining (10) and (12), then setting coefficients of $\delta g^{ij}$ equal
to zero, we arrive at the six field equations

\begin{eqnarray}
   && \hspace{-.3in}
 \frac{c^4}{4\pi G} \left\{\frac{1}{\sqrt{-g}} \frac{\p}{\p x^0}(\sqrt{-g}\,
   g^{00}Q^i_{j0}) - \delta^i_j \frac{1}{\sqrt{-g}} \frac{\p}{\p x^l}
 (\sqrt{-g}\, g^{lm}Q^0_{0m})\right\} \no \\
   && \hspace{1.0in}  
    +\, \sqrt{g^{00}} (T^{\,\,\,i}_j - \delta^i_j T^{\,\,\,0}_0 ) = 0 
\end{eqnarray}
$T^{\mu\nu} $ is the total energy tensor

\beq
      T^{\mu\nu} = T^{\mu\nu}_{\,\,G} + T^{\mu\nu}_{\,\,M}
\eeq
Newton's law of gravitation is to be found, as a first approximation, in
all three diagonal equations.

Before proceeding to the Hamilton equations, we re-express (14) in 
Lagrangian form.  Setting $ \L = \sqrt{-g}\, L $, where 
\beq
      L = L_G + L_M
\eeq
\begin{eqnarray}
  & & \hspace{-.3in}\delta \int \L \, d^4 x = \no \\
  &=& \int d^4 x\biggl\{\frac{\p \L}{\p g_{00}}\delta
 g_{00} + \frac{\p \L}{\p(\p_k g_{00})}\delta (\p_k g_{00}) 
  +\frac{\p \L}{\p g_{ij}} \delta g_{ij} 
  +\frac{\p \L}{\p(\p_0 g_{ij})} \delta (\p_0 g_{ij}) \biggr\} \no \\
 &=& \int d^4 x \biggl\{\left(\frac{\p\L}{\p g_{00}}-\p_k \frac{\p\L}{\p (
 \p_k g_{00})}\right)\delta g_{00} + \left(\frac{\p \L}{\p g_{ij}} - \p_0
 \frac{\p \L}{\p (\p_0 g_{ij})}\right) \delta g_{ij} \biggr\}
\end{eqnarray}
However, $\delta g_{00} = - g_{00} g^{ij} \delta g_{ij} $, therefore

\begin{eqnarray}
 & & \hspace{-.5in} \delta \int \L\, d^4 x = \no \\ 
 & & \hspace{-.4in} = \int d^4 x \biggl\{-g_{00}g^{ij}\left(\frac{\p\L}
 {\p g_{00}}-\p_k\frac{\p\L}{\p(\p_k g_{00})}\right)+\frac{\p\L}{\p g_{ij}}
  - \p_0 \frac{\p \L}{\p(\p_0 g_{ij})}\biggr\} \delta g_{ij}
\end{eqnarray}
In order to satisfy $\disp\delta\int\L\, d^4 x = 0$, the coefficients of
$\delta g_{ij} $ must be zero

\beq
 -g_{00} g^{ij}\left(\frac{\p \L}{\p g_{00}} - \p_k 
  \frac{\p\L}{\p(\p_k g_{00})}\right) + 
  \frac{\p\L}{\p g_{ij}} - \p_0 \frac{\p \L}{\p(\p_0 g_{ij})} = 0 
\eeq
These are identical to field equations (14).

\section*{\large {\bf 2. Hamilton Equations}}

\indent

The six independent dynamical variables $g_{ij}$ possess conjugate momenta
\beq
 \pi^{ij} = \frac{\p\sqrt{-g} \, L_G}{\p(\p_0 g_{ij})} =
            \frac{c^4}{16\pi G} g^{00} g^{ia}
            g^{jb} \p_0 g_{ab} \sqrt{-g}
\eeq
Solving for $\p g_{ij}/\p x^0$ , the Hamiltonian density is 

\begin{eqnarray}
 & & \hspace{-.5in} \H_G = \sqrt{-g}\,H_G = \pi^{ij}\p_0 g_{ij} - \L_G \no\\
 & & \hspace{-.4in} = \frac{8\pi G}{c^4}g_{00}g_{ma}g_{lb}\pi^{la}\pi^{mb}
     \frac{1}{\sqrt{-g}}
  - \frac{c^4}{32\pi G} g^{lm}g^{00}g^{00} 
    \p_l g_{00} \p_m g_{00}\sqrt{-g}
\end{eqnarray}
For simplicity, we now represent matter by the real scalar field

\beq
    L_M = \frac{1}{2}\left(g^{\mu\nu} \p_\mu \phi \, \p_\nu \phi 
           - m^2 \phi\phi\right)
\eeq
with field equations

\beq
   \frac{\p \L_M}{\p \phi} - \p_\mu \frac{\p \L_M}{\p(\p_\mu \phi)} = 0
\eeq
or
\beq
    \frac{1}{\sqrt{-g}}\frac{\p}{\p x^\mu}\left(\sqrt{-g}\, g^{\mu\nu}
    \frac{\p\phi}{\p x^\nu}\right) + m^2 \phi = 0
\eeq
The conjugate momentum is

\beq
 \pi = \frac{\p\sqrt{-g}\, L_M}{\p(\p_0 \phi)} = g^{00} \p_0\phi\,\sqrt{-g}
\eeq  
and Hamiltonian density

\begin{eqnarray}
 \H_M &=& \sqrt{-g}\, H_M = \pi \, \p_0 \phi - \L_M \no \\
      &=& \frac{1}{2} g_{00} \pi \pi \frac{1}{\sqrt{-g}}
    - \frac{1}{2}(g^{lm} \p_l \phi\, \p_m \phi -m^2 \phi\phi)\sqrt{-g} 
\end{eqnarray}

Consider the variation of the spatial integral of $\H = \H_G + \H_M$:

\begin{eqnarray}
 & & \hspace{-.3in} \delta \int \H \,d^3 x = \no \\
 & &= \int d^3 x \biggl\{\frac{\p \H}{\p g_{00}} \delta g_{00} +
 \frac{\p \H}{\p(\p_k g_{00})}
      \delta(\p_k g_{00}) + \frac{\p\H}{\p g_{ij}} \delta g_{ij}
 +\frac{\p\H}{\p\pi^{ij}} \delta \pi^{ij}  \no \\
& &\hspace{.5in}+\frac{\p\H}{\p\phi}\delta\phi
  +\frac{\p\H}{\p(\p_k\phi)}\delta (\p_k \phi)
  + \frac{\p \H}{\p \pi} \delta \pi \biggr \} \no \\
 & &=\int d^3 x \biggl\{\biggl[-g_{00}g^{ij}\left(\frac{\p\H}{\p g_{00}}-
 \p_k\frac{\p\H}{\p(\p_k g_{00})} \right)+\frac{\p\H}{\p g_{ij}}\biggr]
 \delta g_{ij}+\frac{\p\H}{\p\pi^{ij}}\delta\pi^{ij}  \no \\
 & &\hspace{.5in}+\left(\frac{\p\H}{\p\phi} - 
   \p_k\frac{\p\H}{\p(\p_k\phi)}
 \right)\delta \phi + \frac{\p\H}{\p\pi} \delta \pi \biggr\}
\end{eqnarray}
Setting (27) aside for the moment, the definition of $\H$ provides the 
variation

\begin{eqnarray}
 & & \hspace{-.3in} \delta \int \H\, d^3 x = \delta\int d^3 x \left\{
  \pi^{ij} \p_0 g_{ij} + \pi\, \p_0 \phi - \L \right\} \no \\
 & & = \int d^3 x \biggl\{\delta\pi^{ij}\p_0 g_{ij} + \pi^{ij}
  \delta(\p_0 g_{ij}) + \delta \pi\,\p_0\phi + \pi\,\delta(\p_0\phi) \no \\
 & & \hspace{.2in} - \frac{\p \L}{\p g_{00}} \delta g_{00} 
     - \frac{\p \L}{\p(\p_k g_{00})}
 \delta (\p_k g_{00}) - \frac{\p \L}{\p g_{ij}} \delta g_{ij} 
 - \frac{\p \L}{\p(\p_0 g_{ij})} \delta (\p_0 g_{ij}) \no \\
 & & \hspace{.2in}- \frac{\p\L}{\p\phi}\delta\phi - \frac{\p\L}{\p(\p_k \phi)}
  \delta(\p_k\phi) - \frac{\p \L}{\p(\p_0 \phi)} \delta (\p_0 \phi) \biggr\}
\end{eqnarray}
Cancel terms in (28), then integrate by parts, to find

\begin{eqnarray}
 & & \hspace{-.3in} \delta \int \H \, d^3 x = \no \\
 & & = \int d^3 x \biggl\{\biggl[g_{00}g^{ij} \left( \frac{\p \L}{\p g_{00}}
  -\p_k\frac{\p\L}{\p(\p_k g_{00})}\right) -\frac{\p\L}{\p g_{ij}} \biggr]
 \delta g_{ij} + \p_0 g_{ij} \delta \pi^{ij} \no \\
& &\hspace{.6in}-\left(\frac{\p\L}{\p\phi}-\p_k\frac{\p\L}{\p(\p_k\phi)}
  \right) \delta \phi + \p_0 \phi \,\delta \pi \biggr\} 
\end{eqnarray}
Finally, substitute the field equations (19) and (23) 

\beq
 \delta \int \H \, d^3 x = \int d^3 x \left\{-\p_0 \pi^{ij} \delta g_{ij}
 +\p_0 g_{ij}\delta\pi^{ij} -\p_0\pi\,\delta\phi+\p_0\phi\,\delta\pi\right\}
\eeq
The Hamilton equations follow by equating coefficients in (27) and (30):

\begin{eqnarray}
 -\frac{\p \pi^{ij}}{\p x^0} &=& -g_{00}g^{ij}\left(\frac{\p \H}{\p g_{00}}
  - \p_k \frac{\p \H}{\p(\p_k g_{00})}\right) + \frac{\p \H}{\p g_{ij}} \\
 \frac{\p g_{ij}}{\p x^0} & = & \frac{\p \H}{\p \pi^{ij}} \\
 -\frac{\p\pi}{\p x^0}&=&\frac{\p\H}{\p\phi}-\p_k\frac{\p\H}{\p(\p_k\phi)}\\
 \frac{\p \phi}{\p x^0}& =& \frac{\p \H}{\p \pi} 
\end{eqnarray}

\section*{\large {\bf 3. Conservation of Energy }}

\indent

Let us calculate the rate of change of the time-dependent quantity

\beq
    H(x^0) = \int \H \, d^3 x
\eeq
where $\H$ is the total Hamiltonian density:

\begin{eqnarray}
 &&\hspace{-.1in}\frac{dH(x^0)}{dx^0} = \frac{d}{dx^0}\int \H\, d^3 x \no \\
 && = \int d^3 x\biggl\{\frac{\p\H}{\p g_{00}}\p_0 g_{00}
  +\frac{\p\H}{\p(\p_k g_{00})}\p_0(\p_k g_{00}) +\frac{\p\H}{\p g_{ij}}
 \p_0 g_{ij} + \frac{\p\H}{\p\pi^{ij}} \p_0 \pi^{ij} \no \\
 & & \hspace{.5in} + \frac{\p\H}{\p\phi}\p_0\phi  
     + \frac{\p\H}{\p(\p_k\phi)}\p_0(\p_k\phi)
     + \frac{\p\H}{\p \pi} \p_0 \pi \biggr\}\no \\
 && = \int d^3 x \biggl\{
      \biggl[-g_{00}g^{ij}\left(\frac{\p\H}{\p g_{00}}
 - \p_k\frac{\p\H}{\p(\p_k g_{00})}\right) + \frac{\p\H}{\p g_{ij}}\biggr]
 \p_0 g_{ij} + \frac{\p\H}{\p \pi^{ij}}\p_0\pi^{ij} \no \\
 & & \hspace{.5in} + \left(\frac{\p\H}{\p\phi} 
              - \p_k \frac{\p\H}{\p(\p_k\phi)}\right)\p_0\phi
              + \frac{\p\H}{\p\pi} \p_0\pi \biggr \}
\end{eqnarray}
We have made use of 

\beq
 \frac{\p g_{00}}{\p x^0} = -g_{00}g^{ij} \frac{\p g_{ij}}{\p x^0}
\eeq
Here, the volume must be large enough that surface integrals may be 
neglected.  Substitute the Hamilton equations, in order to obtain

\begin{eqnarray}
 \frac{dH(x^0)}{dx^0} &=& \int d^3 x\left\{-\p_0 \pi^{ij}\p_0 g_{ij} + 
 \p_0 g_{ij}\p_0\pi^{ij}-\p_0\pi\,\p_0\phi+\p_0\phi\,\p_0\pi\right\} \no\\
 &=& 0
\end{eqnarray}
Therefore, the integral quantity $H(x^0)$ is conserved, if the field 
equations are satisfied.

The differential law of energy-momentum conservation is [1]

\beq
 \mbox{\rm div}\,T_\mu^{\,\,\,\nu} = \frac{1}{\sqrt{-g}}\frac{\p\sqrt{-g}\,
 T_\mu^{\,\,\,\nu}}{\p x^\nu} - Q^\nu_{\mu\lambda} T_\nu^{\,\,\,\lambda} = 0
\eeq
$ Q^\mu_{\nu\lambda} $ are the connection coefficients of the theory

\beq
 {\nabla}_\nu e_\mu = e_\lambda Q^\lambda_{\mu\nu}
\eeq
(The gravitational field is $Q^\mu_{[\nu\lambda]} = Q^\mu_{\nu\lambda}
- Q^\mu_{\lambda\nu}$.)  Energy conservation is given by

\begin{eqnarray}
 & & \hspace{-.2in} \mbox{\rm div}\,T_0^{\,\,\,\nu} = \frac{1}{\sqrt{-g}}
   \frac{\p \sqrt{-g}\,
 T_0^{\,\,\,\nu}}{\p x^\nu} - Q^\nu_{0\lambda} T_\nu^{\,\,\,\lambda} \no\\
 & & = \frac{1}{\sqrt{-g}}
    \frac{\p\sqrt{-g}\,T_0^{\,\,\,0}}{\p x^0} +
    \frac{1}{\sqrt{-g}}\frac{\p \sqrt{-g}\, T_0^{\,\,\,k}}{\p x^k} -
    Q^0_{00}T_0^{\,\,\,0} - Q^0_{0k}T_0^{\,\,\,k} = 0
\end{eqnarray}
where

\beq
     Q^0_{00} = \frac{1}{2} g^{00}\frac{\p g_{00}}{\p x^0} =
        -\frac{1}{\sqrt{g^{00}}}\frac{\p\sqrt{g^{00}}}{\p x^0}
\eeq
\beq
     Q^0_{0k} = \frac{1}{2} g^{00}\frac{\p g_{00}}{\p x^k} =
        -\frac{1}{\sqrt{g^{00}}}\frac{\p\sqrt{g^{00}}}{\p x^k}
\eeq
Coefficients $Q^j_{00}$ and $Q^j_{0k}$ are identically zero.  It follows 
that

\beq
 \mbox{\rm div}\, T_0^{\,\,\,\nu} = \frac{1}{\sqrt{-h}}
  \frac{\p\sqrt{-h}\,T_0^{\,\,\,0}}{\p x^0}
  + \frac{1}{\sqrt{-h}}\frac{\p\sqrt{-h}\,T_0^{\,\,\,k}}{\p x^k}=0
\eeq
or
\beq
     \frac{\p \sqrt{-h}\, T_0^{\,\,\,0}}{\p x^0} = 
         - \frac{\p \sqrt{-h}\, T_0^{\,\,\,k}}{\p x^k} 
\eeq

On the other hand, the quantity of energy in an infinitesimal region,
$dV_0 = d^3 x$, is given by the first term in the expansion

\begin{eqnarray}
 e_\mu T^{\mu\nu} \sqrt{-g}\,dV_\nu & =& e_0 T^{00} \sqrt{-g}\, dV_0
    + e_0 T^{0k} \sqrt{-g}\, dV_k \no \\
 && +\, {\bfe}_i T^{i0} \sqrt{-g}\, dV_0 + {\bfe}_i T^{ik} \sqrt{-g}\, dV_k
\end{eqnarray}
The scalar basis is a {\it function}, $e_0 = \sqrt{g_{00}}$, and this
is crucial.  It allows us to consider the rate of change of the energy 
integral

\begin{eqnarray}
 & & \hspace{-.5in} \frac{d}{dx^0} \int e_0 T^{00}\sqrt{-g}\, dV_0 = 
   \frac{d}{dx^0}\int 
   \sqrt{g_{00}}\, g^{00} T_0^{\,\,\,0}\sqrt{-g}\, d^3 x \no\\
 & & = \frac{d}{dx^0} \int\sqrt{-h}\, T_0^{\,\,\,0} \, d^3 x = 
    \int \frac{\p \sqrt{-h}\, T_0^{\,\,\,0}}{\p x^0}\, d^3 x 
\end{eqnarray}
This gives zero upon integration of (45) over a sufficiently large volume.
It follows that if (38) is to represent conservation of total energy, then
the integrands in (35) and (47) must be identical

\beq
    \H = \sqrt{-g}\, H = \sqrt{-h}\, T_0^{\,\,\,0}
\eeq
or
\beq
     T_0^{\,\,\,0} = \sqrt{g_{00}}\, H
\eeq
The gravitational Hamiltonian (21) may be evaluated in terms of the 
field $Q^\mu_{[\nu\lambda]}$

\beq
T_{0\,\,G}^{\,\,\,0} = \sqrt{g_{00}}\, H_G = \frac{c^4}{8\pi G}\sqrt{g_{00}}
   \left\{g^{00}Q^l_{m0}Q^m_{l0} - g^{lm}Q^0_{0l}Q^0_{0m}\right\}
\eeq
which implies

\beq
T^{\mu\nu}_{\,\,G} = \frac{c^4}{4\pi G} \sqrt{g_{00}} \left\{
     g^{\mu\alpha}g^{\nu\beta}Q^\rho_{[\eta\alpha]}Q^\eta_{[\rho\beta]}
     - \frac{1}{2} g^{\mu\nu}g^{\alpha\beta}Q^\rho_{[\eta\alpha]}
       Q^\eta_{[\rho\beta]} \right\}
\eeq
In similar fashion, the Hamiltonian (26) gives rise to a factor of
$\sqrt{g_{00}}$ in the matter tensor $T^{\mu\nu}_{\,\,M}$.

\section*{\large {\bf 4. The Poisson Bracket Equation}}

\indent

This section is of a purely formal nature, in which we consider general
dynamical variables, $\U = \sqrt{-g}\, U$, that are functionals of the 
fields, their spatial derivatives, the momenta, their spatial derivatives, 
and the time:

\begin{eqnarray}
 &&\hspace{-.4in}\frac{d U(x^0)}{dx^0} = \frac{d}{dx^0}\int \U\, d^3 x \no\\
 & &\hspace{-.3in} = \int d^3 x \biggl \{\frac{\p\U}{\p g_{00}}\p_0 g_{00}
 + \frac{\p\U}{\p(\p_k g_{00})} \p_0(\p_k g_{00})
 +\frac{\p\U}{\p g_{ij}}\p_0 g_{ij} \no\\
  & &\hspace{-.2in}  + \,\frac{\p\U}{\p(\p_k g_{ij})}\p_0(\p_k g_{ij})
  + \frac{\p\U}{\p\pi^{ij}}\p_0 \pi^{ij} + \frac{\p\U}{\p(\p_k \pi^{ij})}
 \p_0(\p_k \pi^{ij}) \no\\
 & &\hspace{-.2in}  + \,\frac{\p\U}{\p\phi}\p_0\phi
    +\frac{\p\U}{\p(\p_k\phi)}\p_0(\p_k\phi)
    + \frac{\p\U}{\p\pi}\p_0\pi +\frac{\p\U}{\p(\p_k\pi)}\p_0(\p_k\pi)
    + \frac{\p\U}{\p x^0} \biggr \}
\end{eqnarray}
Define the functional derivatives

\begin{eqnarray}
 \frac{\delta\U}{\delta g_{ij}} &=& -g_{00}g^{ij}\left(\frac{\p\U}{\p g_{00}}
  - \p_k \frac{\p\U}{\p(\p_k g_{00})}\right) + \left(\frac{\p\U}{\p g_{ij}}
   - \p_k \frac{\p\U}{\p(\p_k g_{ij})}\right) \\
 \frac{\delta\U}{\delta \pi^{ij}} & =& \frac{\p\U}{\p \pi^{ij}}
           - \p_k \frac{\p\U}{\p(\p_k \pi^{ij})} \\
 \frac{\delta \U}{\delta \phi} &=& \frac{\p\U}{\p\phi} - 
         \p_k\frac{\p\U}{\p(\p_k\phi)} \\
 \frac{\delta \U}{\delta \pi} &=& \frac{\p\U}{\p\pi} - 
         \p_k\frac{\p\U}{\p(\p_k\pi)} 
\end{eqnarray}
In terms of these derivatives, the Hamilton equations (31 -- 34) become

\begin{eqnarray}
  -\frac{\p\pi^{ij}}{\p x^0} &=& \frac{\delta\H}{\delta g_{ij}} \\
   \frac{\p g_{ij}}{\p x^0} &=& \frac{\delta\H}{\delta \pi^{ij}} \\
  -\frac{\p\pi}{\p x^0} &=& \frac{\delta\H}{\delta \phi} \\
   \frac{\p\phi}{\p x^0} &=& \frac{\delta\H}{\delta \pi}
\end{eqnarray}
Integrate (52) by parts and neglect surface terms, in order to obtain
the Poisson bracket equation
\beq
\frac{dU(x^0)}{dx^0} = \int d^3 x \left\{\left(
 \frac{\delta\U}{\delta g_{ij}} \frac{\delta\H}{\delta\pi^{ij}}
 - \frac{\delta\U}{\delta\pi^{ij}} \frac{\delta\H}{\delta g_{ij}}\right)
 + \left(\frac{\delta\U}{\delta\phi} \frac{\delta\H}{\delta\pi}
 -\frac{\delta\U}{\delta\pi} \frac{\delta\H}{\delta\phi} \right)
 + \frac{\p \U}{\p x^0} \right\}
\eeq
Energy conservation (38) is a special case of (61), in which $\H$ is a 
functional of the fields and momenta but is not an explicit function 
of the time.

\vspace{.5in}

\section*{\large {\bf References}}

\begin{enumerate}

\item K. Dalton, ``Electromagnetism and Gravitation,'' {\it Hadronic
   Journal} {\bf 17} (1994) 483; also, http://xxx.lanl.gov/gr-qc/9512027.
\item A. Einstein, ``On the Influence of Gravitation on the Propagation
   of Light,'' in {\it The Principle of Relativity} (Dover, New York, 1952).

\end{enumerate}

\end{document}